\documentclass[useAMS,usenatbib]{mn2e}
\usepackage[T1]{fontenc}
\usepackage{aecompl}
\usepackage{graphicx}
\usepackage{multirow}
\usepackage{amssymb}
\usepackage{hyperref}
\hypersetup{colorlinks=false,
		linkcolor=black,
		urlcolor=cyan}

\title[SN 2012ca]{SN 2012ca: a stripped envelope core-collapse SN interacting with dense circumstellar medium 
\thanks{Based on observations collected at the European Organisation for Astronomical Research in the Southern Hemisphere, Chile, as part of programme 188.D-3003 (PESSTO).} }
\author[C. Inserra et al.]{C. Inserra$^{1}$\thanks{E-mail: c.inserra@qub.ac.uk(CI)},
S. J. Smartt$^{1}$, R. Scalzo$^{2}$, M. Fraser$^1$, A. Pastorello$^3$, M. Childress$^2$,
\newauthor G. Pignata$^4$, A. Jerkstrand$^1$, R. Kotak$^1$, S. Benetti$^3$, M. Della Valle$^{5,6}$, A. Gal-Yam$^7$,
\newauthor P. Mazzali$^{8,3,9}$, K. Smith$^1$, M. Sullivan$^{10}$, S. Valenti$^{11,12}$, O. Yaron$^7$, D. Young$^1$
\newauthor and D. Reichart$^{13}$\\
$^{1}$Astrophysics Research Centre, School of Mathematics and Physics, Queen's University
  Belfast, Belfast BT7 1NN, UK\\
$^{2}$Research School of Astronomy and Astrophysics, The Australian National University, Weston Creek, ACT 2611, Australia\\
$^{3}$INAF - Osservatorio Astronomico di Padova, Vicolo dell'Osservatorio 5, I-35122 Padova, Italy\\
$^{4}$Departamento de Ciencias Fisicas, Universidad Andres Bello, Avda. Republica 252, Santiago, Chile\\
$^{5}$INAF - Osservatorio astronomico di Capodimonte, Salita Moiariello 16, I- 80131 Napoli, Italy\\
$^{6}$International Centre for Relativistic Astrophysics, Piazzale della Repubblica 2, 65122 Pescara, Italy\\
$^{7}$Benoziyo Center for Astrophysics, Weizmann Institute of Science, 76100 Rehovot, Israel\\
$^{8}$Astrophysics Research Institute, Liverpool John Moores University ,Liverpool, UK\\
$^{9}$Max-Planck Institut f\"ur Astrophysik, Karl-Schwarzschildstr. 1, D-85748 Garching, Germany\\
$^{10}$School of Physics and Astronomy, University of Southampton, Southampton, SO17 1BJ, UK\\
$^{11}$Las Cumbres Observatory Global Telescope Network, 6740 Cortona Dr, Suite 102, Goleta, CA 93117, USA\\
$^{12}$Department of Physics, University of California, Santa Barbara, Broida Hall, Mail Code 9530, Santa Barbara, CA 93106-9530, USA\\
$^{13}$University of North Carolina at Chapel Hill, Campus Box 3255, Chapel Hill, NC 27599-3255, USA}

\def\kms{km\,s$^{-1}$}
\def\Ha{H{$\alpha$}}
\def\Hb{H{$\beta$}}

\def\co{$^{56}$Co}

\def\ca{SN2012ca}

\begin{document}

\date{Received.....; accepted...........}

\pagerange{\pageref{firstpage}--\pageref{lastpage}} \pubyear{}

\maketitle

\label{firstpage}

\begin{abstract}

  We report optical and near-infrared  observations of 
  SN2012ca with the Public ESO Spectroscopy Survey of Transient Objects (PESSTO), spread over one year since 
  discovery. The supernova (SN) bears many similarities to SN1997cy and 
to other events classified as Type IIn but which have been 
suggested to have a thermonuclear origin with narrow hydrogen
lines produced when the ejecta impact a hydrogen-rich circumstellar
medium (CSM). Our analysis, especially in the nebular
  phase, reveals the presence of oxygen, magnesium and carbon 
  features. 
  This suggests a core collapse explanation for \ca, in
  contrast to the thermonuclear interpretation proposed for some members of this group.
We  suggest that the data can be explained with a hydrogen and helium
deficient SN ejecta (Type I) interacting with a hydrogen-rich CSM, 
but that the explosion was more likely a Type Ic core-collapse explosion
than a Type Ia thermonuclear one. This suggests 
 two channels (both thermonuclear and stripped envelope core-collapse) may be responsible for these SN~1997cy-like 
events.


 \end{abstract}
 

\section{Introduction}\label{sec:intro}

Supernovae (SNe) are produced by two physical mechanisms:
thermonuclear SNe (SN Ia) which completely destroy the degenerate
progenitor star \citep{hn00}, and core-collapse SNe (CC-SNe) which
leave a compact remnant \citep{2012ARNPS..62..407J}. Thermonuclear SNe
can be produced through a single-degenerate channel when a white
dwarf accretes hydrogen and helium from a companion or through the
merger of two WDs \citep{hn00}. CC-SNe are produced by the explosion following the
gravitational collapse of the cores of massive stars and classified
according to the presence (SNe II) or absence (SN I) of H and/or He in
their spectra \citep{fi97}. Theoretical stellar evolution calculations
have long attempted to link evolved model stars
\citep{he03,2012ARA&A..50..107L} to the types of CC-SNe observed and
in some cases the directly identified progenitors
\citep{2009ARA&A..47...63S}. The two physical mechanisms are rather
different, but a number of SNe have been discovered for which 
 the underlying origin of the explosion is unclear. 

Events such as SN~2002ic \citep{ha03,de04,wv04} and PTF11kx
\citep{di12,si13a}, exhibit unambiguous signs of the ejecta
interacting with circumstellar material (CSM). They have been
classified as interacting Type Ia SNe and are thought to be of
thermonuclear origin. Their spectra appear to be a``diluted'' spectrum
of a bright SN Ia (e.g. SNe 1991T and 1999aa) along with superimposed  H
emission lines. PTF11kx shows the strongest evidence
for being a thermonuclear event interacting with CSM expelled by a
companion red giant star \citep{di12}. 
However the physical
origin of other SNe are still uncertain and this debate dates back to
the peculiar Type IIn SN1997cy \citep{ge00,tu00}. The growing
arguments that many such events with narrow hydrogen (classified as
SNe IIn) could be thermonuclear has resulted in SN2008J \citep{ta12}
being labelled as an interacting SN Ia, while new data
for SN2005gj \citep{al06,pr07} has been used to suggest a 
thermonuclear origin rather than a CC-SN \citep{si13b}.  But this is not
the unanimous view, with \citet{be06} and \citet{tr08} 
arguing for a core-collapse origin for SNe 2002ic and 2005gj, respectively. 

These ambiguous and interacting events are important for
determining the possible progenitor channels for SNe Ia in particular,
but are rare. During the first run of the Public ESO Spectroscopy Survey of Transient Objects (PESSTO)\footnote{www.pessto.org} in
April 2012, SN2012ca was classified as an unusual Type IIn, with an
early spectral resemblance to SN~1997cy. Its distance and brightness
have allowed an extensive follow-up until late epochs providing
another object to better understand whether these events are all
interacting SNe Ia or some are in fact CC-SNe. The letter presents
most of the first season of PESSTO optical and NIR data, which is
publicly available in the ESO\footnote{as part of the Spectroscopic Survey Data Release 1 
(SSDR1)} and WISeREP\footnote{http://www.weizmann.ac.il/astrophysics/wiserep/} \citep{ya12} archives. 


\section{Observations and data analysis}\label{sec:obs}

\ca\/ was discovered in the late-type spiral galaxy ESO 336-G009 by
\citet{dr12}, with a first detection on 2012 April 25.6 UT
(m$_r\sim$14.8 mag). A spectrum was obtained at the New Technology
Telescope (NTT) + EFOSC2 \citep{va12,in12} on April 29.4 as part of
PESSTO. The initial spectrum of \ca\/ showed a resemblance to a few
SNe 1997cy-like explosions at $\sim$60 d post maximum. Because of the lack of
early data, we cross-correlated the first spectrum with a library of
SN 1997cy-like events at multiple epochs
and found a best match adopting a peak light epoch at  MJD $55988.0\pm9.0$ (March 2).
From the emission
component of the Balmer lines a redshift $z=0.019$ was measured,
consistent with that of the host galaxy. Adopting a standard cosmology
with $H_0 = 72$ \kms\/, $\Omega_{\rm M}=0.27$ and $\Omega_{\rm
  \lambda}=0.73$, NED\footnote{NASA/IPAC Extragalactic Database}
provides $\mu=34.54\pm0.15$ mag from the heliocentric radial velocity of
$v_{\rm host}=5834\pm37$ \kms\/, which will be used throughout the
letter. There is no detection of Na~{\sc i} interstellar medium (ISM)
features from the host galaxy, nor do we have any evidence of
significant extinction inside the host from the SN spectra itself.
This suggests that the internal absorption  is low and we assume it to be  
negligible. Moreover, \ca\/ is 3.5 kpc far from the centre of the host galaxy, 
this is consistent with the absence of significant extinction. The Galactic reddening
toward the SN line of sight is $E(B-V)=0.06$ mag \citep{sf11}, which
we correct for in the following.
Spectro-photometric follow-up was obtained by PESSTO with NTT+EFOSC2, PROMPT \citep[][]{re05}, Swift+UVOT and ANU+WiFeS in optical, while in near infrared (NIR) with NTT+SOFI.
Images and spectra
were reduced in the standard fashion, and the NTT data 
were processed within the PESSTO pipeline as  in \citet{va13,fr13}.
The resolution of the optical spectra were checked and found to be
$\sim$18 \AA\/ and $\sim$2 \AA\/ for the EFOSC2 and WiFeS data,
respectively. The resolution in the NIR were $\sim$23 \AA\/ (blue grism) and $\sim$33 \AA\/ (red grism). 

\subsection{Bolometric luminosity}\label{ss:bol}
\begin{figure}
\includegraphics[width=\columnwidth,height=5.7cm]{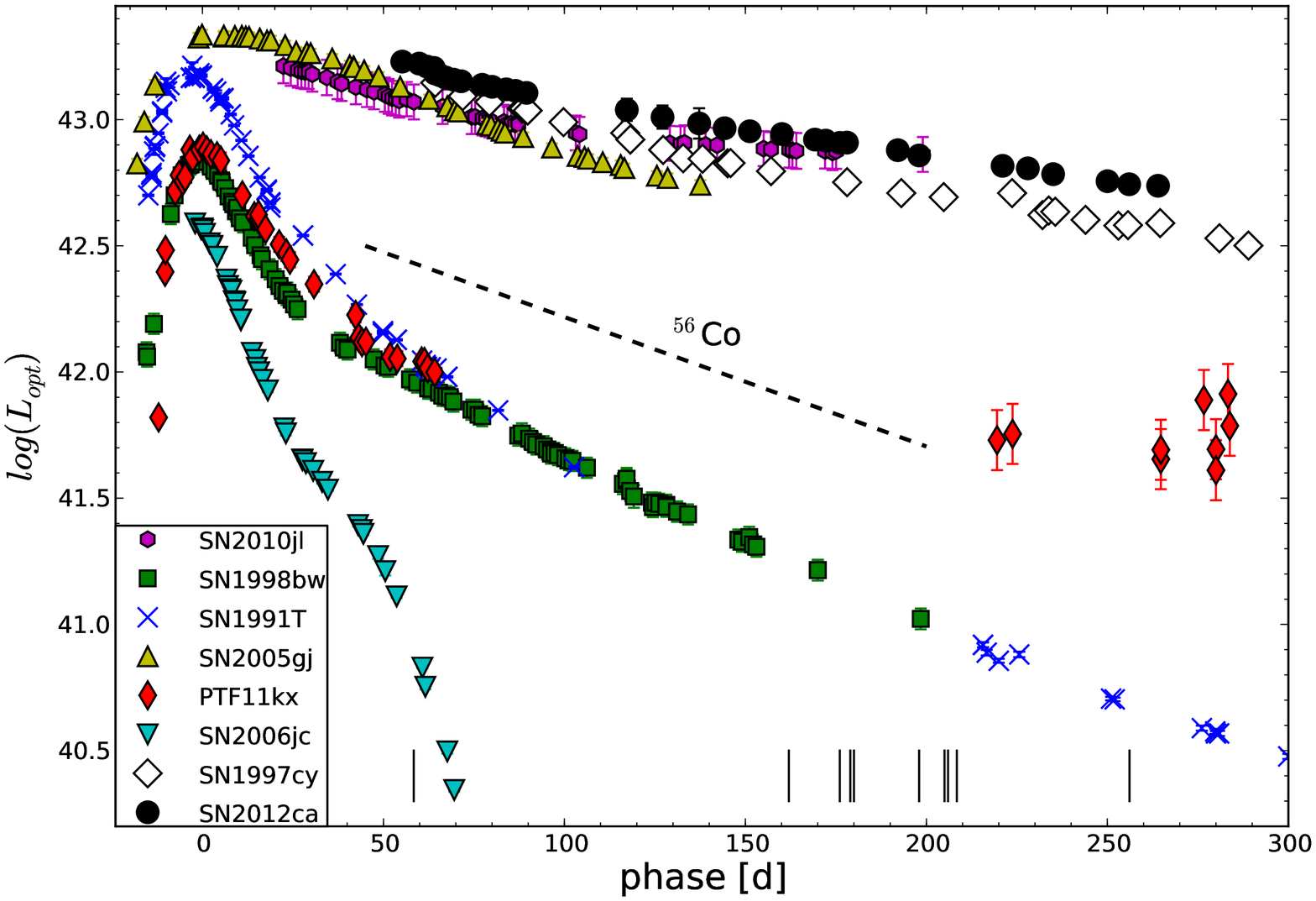}
\caption{$Ugriz$ (optical) bolometric light curves of \ca\/, the Type
  Ic SN 1998bw \citep[{\it
    UBVRI};][]{ga98} 
  , the bright Type Ia 1991T \citep[{\it UBVRI};][]{li98}, the
  Type Ia interacting PTF11kx \citep[$gri$;][]{di12,si13a} and SN
  2005gj \citep[{\it ugriz};][]{pr07}, the Type IIn 2010jl
  \citep[$UBVRI$;][]{zh12}, the Type Ibn SN 2006jc \citep[{\it
    UBVRI};][]{pas07} and the peculiar Type IIn SN 1997cy
  \citep[$BVRI$;][]{ge00,tu00}. 
  Phase is with respect to maximum light
  at rest-frame. The epochs of the SN~2012ca spectra reported are marked with
  vertical lines.
}
\label{fig:cfr_bol}
\end{figure}

\begin{figure}
\includegraphics[width=\columnwidth,height=6.0cm]{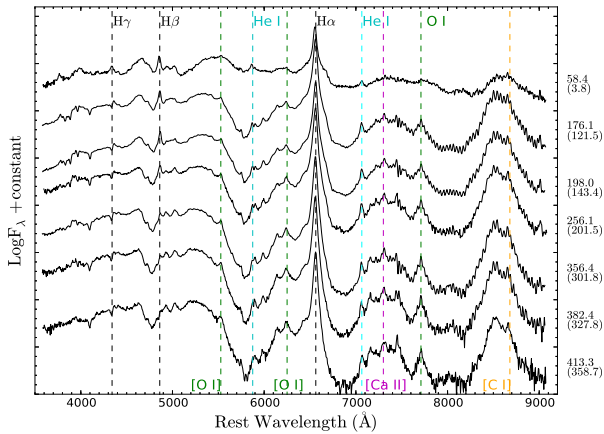}
\caption{Rest-frame sequence of spectra of \ca\/ until one year since discovery. 
Phases are since the estimated maximum epoch (phases since discovery are reported in parentheses). }
\label{fig:spev}
\end{figure}


A pseudo-bolometric light curve ($Ugriz$) of \ca\/ (Fig.~\ref{fig:cfr_bol}) was constructed
using similar methods as in \cite{in13}.
There is certainly flux outside these bands however, to first order we
assume that these contributions are similar in each SN. This allows
an approximate comparison of the pseudo-bolometric luminosities which
is not as comprehensive as full bolometric measurements, but better
than using single band comparisons.
Although there are uncertainties in the explosion date, peak epoch
and some caveats comparing pseudo-bolometric lightcurves determined using different optical filter systems,
\ca\/ appears brighter and more slowly declining than SN2005gj. It is
of comparable luminosity to SN1997cy between 60-250 d from peak, 
and also to the luminous Type IIn SN2010jl.
However it is significantly brighter and slower evolving than SNe 1991T (Ia) and 1998bw (Ic)
or the interacting Ibn SN2006jc. Indeed, \ca\/ appears to be 10
times more luminous than the bright SN Ic reported in Fig.~\ref{fig:cfr_bol},
and its overall evolution is much slower than that of normal Type Ic or Type Ia 
events. 
The post peak decline rate is $\gamma=0.5\pm0.1$ mag/100 d until the last
epoch available, and is much slower than those of the interacting SNe Ia
PTF11kx and SN2005gj and the Type Ibn 2006jc, with $\gamma\sim3.1$,
$\gamma\sim1.1$ and $\gamma\sim7.4$ mag/100d, respectively. The
decrease is slower than the \co\/ decay even at 250 d, suggesting that
the interaction is still the main energy source. One striking feature in Fig.~\ref{fig:cfr_bol}
is the self-similarity of \ca\/, SN1997cy, SN2005gj and how different they
appear compared to the initial decline of PTF11kx. While the PTF11kx decline 
$\gamma\sim0.5$ after 200 d is similar to that of \ca\/, the late-time values of the former
 are determined solely from $r$-band observations (assuming constant colours from neighbouring epochs) and show considerable scatter, 
 thus we suggest to treat this result with caution.

\subsection{Spectroscopy}\label{ss:sp}

\begin{figure*}
\includegraphics[width=18cm,height=5.7cm]{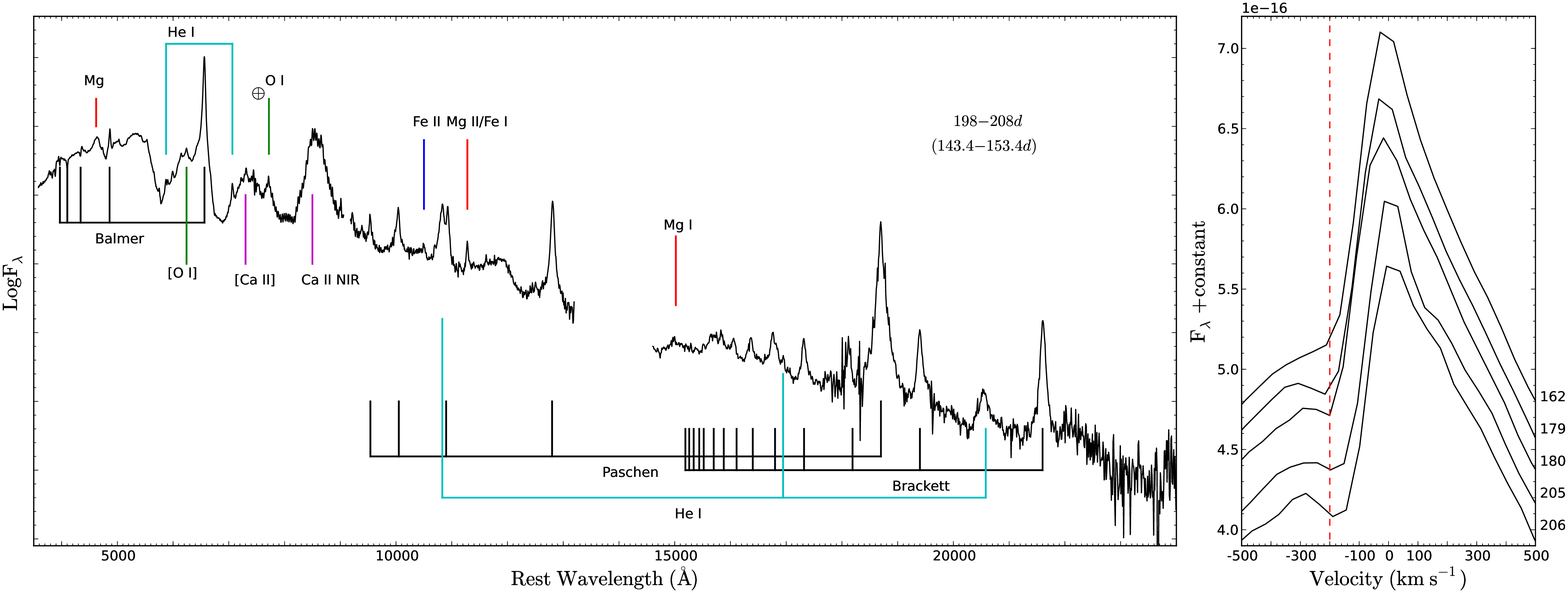}
\caption{Right: composite spectrum of \ca\/ from optical to NIR wavelengths, at $\sim$198-208 d post maximum ($\sim$143-153 d from discovery). 
The $\oplus$ symbol marks the position of the strongest telluric absorption. The most prominent features are labelled. Right: temporal evolution of \Ha\/ from  162 d to 206 d post maximum. The vertical red dashed line marks the velocity of the CSM.} 
\label{fig:sptot}
\end{figure*}

Optical spectra of \ca\/ are shown in Fig.~\ref{fig:spev} to
illustrate the evolution during the first year. The somewhat blue
continuum does not show significant evolution, suggesting an
origin due to multiple, narrow lines of iron-group elements creating a
pseudo continuum \citep[][]{sm09,si13b}. After about 180 d the
pseudo-continuum seems visible only at wavelength shorter than
5500 \AA\/, whereas the features redward look nebular. Most notable in
\ca\/ is the presence of Balmer and He~{\sc i} emission lines through
its entire evolution. The high resolution of WiFeS allows us to
resolve narrow components in H lines. We see narrow P-Cygni absorption
in \Ha\/ from a wind, as shown in Fig.~\ref{fig:sptot} (right). The
expansion velocity of the absorption minimum is $200\pm90$ \kms\/ , a
factor 3 greater than that measured for PTF11kx \citep[65
\kms;][]{di12}. This is also a factor 2 or 3 greater than that observed in
probable diffuse CSM from the companion of an exploding WD in some
SNe Ia \citep[50-100 \kms;][]{pa11}. However, \citet{st11} noted that some higher velocities
(150-200 \kms) can be reached in the CSM of a small fraction of SNe Ia.
A broader emission component is also identified for H and He~{\sc i}.
No P-Cygni profiles linked to this component are seen. 
The He~{\sc i} broader component
shows a FWHM velocity $\sim1800$ \kms\/, decreasing to $\sim1500$
\kms\/ in the last spectrum, while \Ha\/ has $v_{\rm
  FWHM}\sim3000-2200$ \kms\/ in the same period. This is also
confirmed by the measurement of the unblended NIR features (see
below). These broad lines can be associated with the fast-moving
ejecta or, in case of interaction such as in \ca\/, viable
explanations are boxy emission from interaction \citep{cf94} or
Thomson scattering of photons by electron in a dense CSM
\citep{ch01,de09}. From deblending  the Ca~{\sc ii} NIR
components, we measure a velocity of $\sim7000$ \kms\/ (at +198d)
related to the fast-moving ejecta, higher than those of H and He,
suggesting a different origin for those emission components.

A combined spectrum comprised of an optical and a blue NIR grism taken on Sept. 15, plus a red NIR grism taken on Sept. 24 is shown in Fig.~\ref{fig:sptot}.
As reported above, the optical region shows Balmer lines in emission, while the Paschen and Brackett series are visible at wavelengths $>8500$ \AA\/ with comparable velocities to the Balmer series at that phase ($v{\rm (Pa\alpha)}_{\rm FWHM}=v{\rm (H\alpha)}_{\rm FWHM}\sim2400$ \kms). He~{\sc i} $\lambda$5876, $\lambda$7065, $\lambda$10830 and $\lambda$20589 emission lines are clearly detected, and also tentatively at $\sim16800$~\AA\/. Calcium has also been identified with a weak Ca~{\sc ii} H\&K line, the Ca NIR triplet and the [Ca~{\sc ii}] $\lambda\lambda7291,7324$ doublet, implying that at 198d the SN is moving toward the nebular phase. 
Two puzzling lines at $\sim6250$~\AA\/ and $\sim7710$~\AA\/ are detected. They could be [O~{\sc i}] $\lambda\lambda$6300, 6364 and O~{\sc i} $\lambda$7774 (v$_{\rm FWHM}\sim7000$ \kms), but blue shifted by $\sim2500$~\kms\/. This hypothesis is strengthened by the presence of a line at $\sim7170$ \AA\/ identifiable as [O~{\sc ii}] blue shifted at the same velocity. The shift could be attributed to asymmetries in the ejecta or a clump moving toward us. Moreover, [O {\sc i}] $\lambda$5577 blue shifted by the same amount could be responsible for the red shoulder of the pseudo continuum.
Metal lines of various species are unambiguously detected both in the optical and 
the NIR. Fe~{\sc ii} multiplet~42 
is responsible for the features redder than \Hb\/.  Mg~{\sc i}] $\lambda$4571 is already seen at this phase and gives plausibility to the detection of [O~{\sc i}] because of their mutual presence in Type Ib/c SNe \citep[see][]{hu09}. Other Mg lines could explain the feature at $\sim4600$ \AA\/ in the first spectrum and at 11300 \AA\/ at $\sim198$d, whilst Mg~{\sc i} $\lambda$15024, the strongest Mg line in the NIR, is detected and we measured a FWHM velocity of $\sim$7000 \kms\/. 
After 256~d the Ca NIR triplet is weaker, and [C~{\sc i}] at $\sim8700$~\AA\/ is shown in emission. The [C~{\sc i}] profile resembles 
those of typical SNe Ic after five months from peak \citep{fi97}.

\begin{figure}
\includegraphics[width=\columnwidth,height=6.9cm]{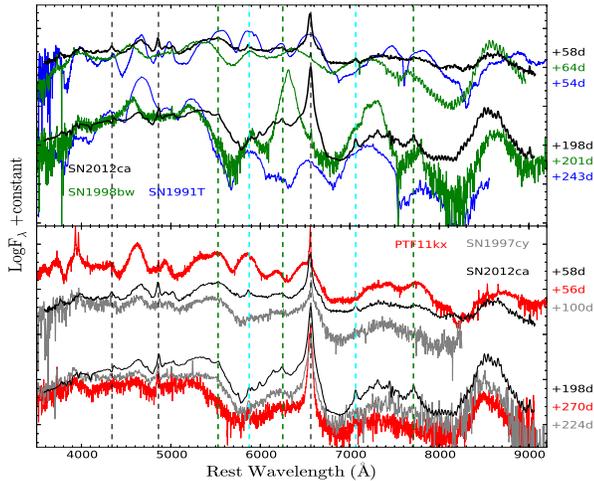}
\caption{Top: comparison of two spectra of \ca\/ with those of SNe 1991T \citep[Ia,][]{gl98} and 1998bw \citep[Ic,][]{pa01}. Bottom: comparison of the  two spectra of \ca\/ with those of PTF11kx \citep[][]{di12} and SN1997cy \citep[][]{tu00}. Balmer, He~{\sc i}, O~{\sc i} and [O~{\sc i}] features are indicated by vertical black, cyan and green dashed lines, respectively. 
Phases with respect to maximum light.} 
\label{fig:spcmp}
\end{figure}

Since SNe Ia and Ic have similar spectral evolution during the
photospheric period and as a consequence it is not easy to detect
features uniquely seen in a class before $\sim$60~d post maximum
\citep{ta06}, especially if we take into account the interaction. In Fig.~\ref{fig:spcmp} we compare \ca\/ 
spectra
with those of the bright  SNe 1998bw (Ic) and 1991T (Ia) at
similar epochs to investigate possible spectral similitudes.
In the earliest spectrum of \ca\/ the blue region ($\lesssim5600$ \AA\/) 
appears more similar to SN~1998bw than SN~1991T both in continuum and
line profiles. In the region around \Ha\/ we cannot make a detailed
comparison because of the presence of the emission lines, however the
Si~{\sc ii} and [Fe~{\sc ii}] 
lines typical of Type Ia SNe are not consistent with the line profiles
seen in \ca\/, even after accounting for dilution by a
continuum. 
In the nebular phase comparison, a SN Ia spectrum at a similar phase
to \ca\/ makes it difficult to explain the region bluer than 6500\AA\/
because of the presence of [Fe~{\sc ii}], [Fe~{\sc iii}] and [Co~{\sc
  ii}] lines in the
former. 
In contrast, \ca\/ bears more resemblance to
  SN1998bw. We suggest the emission features at 6250\AA\/ and 7710\AA\ are
  indeed [O~{\sc i}] which would make a Ia origin unlikely. While the
  features are not as broad and prominent as in SN1998bw, we suggest
  they are most likely to be [O~{\sc i}], and hence are important
  indications of a core-collapse origin. 
 In the bottom panel of Fig.~\ref{fig:spcmp} the
  difference between an interacting Type Ia and \ca\/ becomes clear,
  comparing the object to PTF11kx at similar
  epochs. 
  They show some 
  dissimilarities, 
  mainly due
  to the presence of strong metal lines such as Fe~{\sc ii} 
  in
  PTF11kx. 
  The 198 d spectrum is more similar to that of SN1997cy at $\sim$224 d
  \citep[epoch defined in][]{tu00} and both show a spectral evolution in the [O~{\sc i}] lines regions different to
  that of PTF11kx. 
  \ca\/ and SN 1997cy have similar He~{\sc i} emission lines 
  (EW$_{{\rm 12ca}}^{\lambda5876}=$ EW$_{{\rm 97cy}}^{\lambda5876}\sim7$\AA\/, EW$_{{\rm 12ca}}^{\lambda7065}\sim16$\AA\/ and EW$_{{\rm 97cy}}^{\lambda7065}\sim18$\AA\/), and \Ha\/ FWHM velocity ($v_{\rm 12ca}\sim2900$ \kms\/ and $v_{\rm 97cy}\sim3200$ \kms\/). 


\section{Is \ca\/ an interacting CC-SNe?}\label{sec:dis}

The discovery of PTF11kx and its nature has changed the paradigm of
these SN1997cy-like events, leading \citet{si13b} to suggest a
thermonuclear origin for the majority, or even all, of these type of events. 
The data for SN2012ca indicate that it too is similar to SN1997cy and we consider it more likely to have a core-collapse origin.  The 
evidence for a core collapse interpretation are summarised as follows: 

\begin{itemize}

\item The overall spectral evolution of \ca\/ is hard to explain using
  diluted Type Ia spectra with superimposed emission. SNe Ia 
 over one year
  are dominated by forbidden iron lines, as also shown by \citet{fi92}
  for the luminous SN1991T. On the contrary, a broad-line
  Type Ic such as SN1998bw shows more spectral similarities.

\item O~{\sc i}, [O~{\sc i}] and [O~{\sc ii}] lines were likely
  identified, blue shifted by $\sim$2500 \kms\/.  A number of magnesium lines were
  identified, including Mg~{\sc i} $\lambda$15024 which is usually
  seen in spectra of Type Ic SNe 
  \citep[e.g. SNe 1998bw,
  2004aw;][]{pa01,ta06}. Both elements, and their observed line strengths are more
indicative of a massive progenitor and  CC-SN origin. 

\item  We detect [C~{\sc i}] at 8700 \AA\/ at 413 d
  with $v_{\rm FWHM}\gtrsim 1500$~\kms\/. This is again similar to
  what is typically seen in the ejecta of Type Ic SNe
  \citep[e.g.][]{fi97}.

\item The Ca~{\sc ii} NIR triplet is usually seen during the
  photospheric phase of Type Ia SNe \citep{fi97}. 
  Forbidden [Ca~{\sc ii}] 
  has been seen only in sub luminous SNe Ia \citep{fi92a}, but not in normal or bright (e.g. SN 1991T) SNe Ia.
  The SNe Ia ejecta environment is typically more strongly ionised
  than in SNe Ic, thus the absence of [Ca~{\sc ii}] in Type Ia SNe is
  probably linked to the fact that it is at a higher ionisation state \citep{li97,ma10}.
  Any possible interaction should enhance this effect, and so the
  presence of [Ca~{\sc ii}] in \ca\/ again points towards core
  collapse.

\item \Ha\/ narrow P-Cygni absorption from a wind has been detected
  with an expansion velocity of $\sim$200 \kms, a factor $\sim3$
  greater than what was seen in PTF11kx.

\item Broader and persistent He~{\sc i} emission lines have been detected in \ca\/, in contrast to what seen in PTF11kx.
Furthermore, it appears that little to no He~{\sc i} emission is a key attribute of members of the SN Ia-CSM class \citep{si13b}.


\end{itemize}

The data favour the interpretation of \ca\/ as a CC-SN, 
with a certain degree of asymmetry in the ejecta and surrounded by a H- and He-rich CSM. The interacting SN Ia scenario 
appears to work well for for PTF11kx but the link has been extended broadly to many SNe showing interaction
which are SN1997cy-like. We argue that this is almost certainly premature, and that the extensive data
set for SN2012ca, including the bolometric lightcurve and optical to NIR spectral sequence 
is strongly suggestive of a core-collapse origin. 
The similarity to bright SNe Ic nebular spectra at phases of $>$150d
suggests that it could well have been a stripped envelope progenitor
which impacts and interacts with an H-rich CSM. 
Although the 
presence of H in the spectra and narrow emission lines  suggest a 
classification as a Type IIn SN by definition, we do not find clear evidence that the SN 
ejecta was H-rich.  
This illustrates the limitations of our currently employed SN classification
nomenclature. Indeed SN2012ca and probably SN1997cy are better referred to as 
interacting stripped envelope SNe. 
\ca\/ provides evidence of a viable alternative scenario to the
thermonuclear interpretation \citep[Ia-CSM,][]{si13b} about interacting SNe with H and He rich
CSM, suggesting that two different channels may be responsible for SN1997cy-like events.

\section*{Acknowledgments}
Based on observations collected by PESSTO, 
and the Panchromatic Robotic Optical Monitoring and Polarimetry
Telescope (PROMPT), Chile; 
the Australian National
University 2.3m Telescope and the Swift Satellite.
Funded by the European Research
Council under the European Union's Seventh Framework Programme
(FP7/2007-2013)/ERC Grant agreement n$^{\rm o}$ [291222] (SJS). SB and AP  acknowledge the PRIN-INAF 2011 project ``Transient Universe: from ESO Large to PESSTO". 
G.P.  acknowledges  Millennium Center for Supernova Science (P10-064-F).

\end{document}